\documentclass{aa}
\usepackage{graphicx}
\usepackage{natbib}
\usepackage{color}
\bibpunct{(}{)}{;}{a}{}{,} 
\begin{document}

\def\ms{M$_{\odot}$}
\def\zs{Z$_{\odot}$}

\title{On the self-enrichment scenario of galactic globular clusters: Constraints 
on the IMF}

\author{ Nikos Prantzos\inst{1} and Corinne Charbonnel\inst{2,3}
        }

\authorrunning{N. Prantzos and C. Charbonnel}
 
\titlerunning{Self-enrichment and the IMF of globular clusters}


\institute{ Institut d'Astrophysique de Paris, UMR7095 CNRS, Univ.P. \& M.Curie,  98bis Bd. Arago, 75104 Paris, France, 
                \email{prantzos@iap.fr}
\and 
Geneva Observatory, CH 1290 Sauverny, Switzerland
\email{Corinne.Charbonnel@obs.unige.ch}
\and
LATT UMR 5572 CNRS, 14, av.E.Belin, 31400 Toulouse, France
           }
\date{Submitted April 5, 2006}

\abstract{}{Galactic globular cluster (GC) stars exhibit abundance 
patterns which are not shared by their field counterparts, e.g.  the well documented O-Na and  Mg-Al anticorrelations. 
A widely held  hypothesis is that the gas was 
``polluted" by stars more massive than the presently observed low-mass stars. 
In the framework of this
``self-enrichment" scenario for GCs, we 
present a new method to derive the Initial Mass Function (IMF) of the polluters, 
by using the O/Na abundance distribution.}
{We focus on NGC~2808, a GC for which the largest sample of O and Na abundance determinations  is presently available. We use the abundance distribution of [O/Na] to derive the amount of polluted material with respect to that of original composition.
We explore in details two scenarios for the self-enrichment of the cluster, which differ by the assumptions made on the composition of the polluter ejecta. In each case 
we consider two classes of possible ``culprits" : massive Asymptotic Giant Branch (AGB) stars (4-9~\ms)  and winds of massive stars (WMS) in the mass range 10-100~\ms.}
{We obtain upper limits for the slope of the IMF (assumed to be given by a power-law) of the stars initially more massive than the present turnoff mass. We also derive lower limits for the amount of stellar residues in NGC~2808.}
{We find that the polluter IMF had to be much flatter than presently observed IMFs in  stellar clusters; this is  in agreement with the results of two other methods for GC IMF determination, which we also discuss.
Additionaly, we find that the present mass of the GC should be 
totally dominated by stellar remnants if the polluters were AGB stars; this is 
not the case if the culprits are WMS. We critically analyse the advantages and shortcomings of each potential polluter class, and we find the WMS scenario more attractive.}


\maketitle

\section{Abundance anomalies in galactic globular clusters and the self-enrichment scenario}
Globular cluster (GC) studies are linked to a large variety of astrophysical issues.
These stellar systems provide indeed insights into the formation and early chemical evolution
of galaxies and of their substructures. They also represent unique laboratories
for dynamical processes that occur on timescales shorter 
than the Hubble time (such as two-body relaxation, mass segregation from equipartition
of energy, stellar encounters, collisions and mergers, core collapse, tidal stripping). 
Because they are large aggregates of coeval stars, all located at the same distance from us,
and which contain both very common and unusual (e.g., blue stragglers) stellar objects, they provide invaluable tests to stellar evolution theory. However, and despite their broad astrophysical importance, their origin and formation processes still remain uncertain.

For all these reasons 
galactic GCs have been subject to intensive scrunity over the past decades.
The advent of very large telescopes and of multiple-object spectrometers allowed  
a detailed investigation of their chemical composition 
(see Gratton et al. 2004 and Sneden 2005 for recent and extended reviews). 
We briefly recall here the fundamental chemical properties of the galactic GCs : 

(i) Most of them  appear to be
mono-metallic as far as the Fe-group elements (Fe, Ni, Cu) are concerned
(with the notable exception of Omega Cen);

(ii) They present very low scatter and the same trends as field stars
for the neutron-capture elements (Ba, La, Eu) and the alpha-elements (Si, Ca);

(iii) For the lighter elements (from C to Al) they exhibit  complex patterns 
and large star-to-star abundance variations, which are not shared by their field counterparts.
Among these anomalous patterns, the most striking ones are the anticorrelations between
the abundances of O and Na (see Pilachowski 1984, Drake et al. 1992 and Sneden et al. 1991
for the early studies; for complete references see Gratton et al. 2004) and 
of Mg and Al (see e.g. Ivans et al. 1999, Ramirez \& Cohen 2002)\footnote{Star-to-star 
variations of the strengths of the lines of these elements (Na and Al in particular) 
were known to exist at the tip of the red giant branch (RGB) for a long time 
(Cohen 1978; Peterson 1980; Norris et al. 1981; see Kraft 1994 for an early review).}.

These anticorrelations are easily understood as the result
of proton-capture nucleosynthesis in the CNO-cycle and the NeNa- and MgAl-chains
of H-burning (Denissenkov \& Denissenkova 1990; Langer et al. 1993). 
Proton-captures on $^{16}$O and $^{22}$Ne respectively lead to the destruction 
of O and to the production of $^{23}$Na at  temperatures $> 2 \times 10^7$~K.  
At higher temperatures ($\sim 4 \times 10^7$~K), 
$^{20}$Ne generates $^{23}$Na. 
On the other hand the MgAl-chain is only active at temperatures higher 
than $\sim 7 \times 10^7$~K. 

Although the underlying nucleosynthesis is unambiguous, the identification of the 
corresponding astrophysical site still remains a challenge.
For a long time the main question was to know if such patterns were inherited at the 
birth of the stars that we are currently observing (the so-called {\it self-enrichment hypothesis}) 
or if they were generated in the course of the evolution of these objects 
(the so-called {\it evolution hypothesis}).
A major breakthrough was made in that respect from high resolution spectrographs on 
very large telescopes that allowed one to determine the abundances of the 
concerned elements in faint subgiant and turnoff stars in a couple of GCs 
(Gratton et al. 2001; Grundhal et al. 2002; 
Carretta et al. 2003, 2004; Ramirez \& Cohen 2002, 2003).
The discovery of the above-mentioned anticorrelations in these scarcely evolved objects 
has given a new spin to the self-enrichment scenario.
Indeed turnoff stars are not hot enough for the required nuclear reactions to occur 
in their interiors.  
Additionally the spread of abundances cannot be explained by in situ mechanisms 
like gravitational setlling and thermal diffusion (Richard et al. 2002, 
Michaud et al. 2004). 
This implies that the O, Na, Mg and Al abundance anomalies pre-existed 
in the material out of which these objects formed; their gas must have been polluted 
early in the history of the cluster by more  massive and faster evolving 
stars (Cottrell \& Da Costa 1981)\footnote{An alternative would be that the abundance 
anomalies were accreted onto the surface of low-mass stars from the mass loss by massive 
stars (D'Antona et al. 1983; Thoul et al. 2002). In this case only the surface convective 
layers would have a polluted composition, and the anomalies would disappear on the red 
giant branch due to dilution during the first dredge-up event. 
However turnoff and bright giant stars 
exhibit the same anti-correlations; this means that the entire star, or at least 80$\%$ 
of its mass (which corresponds approximately to the maximum depth of the convective envelope 
during the first dredge-up) has the composition of the photosphere.}. 

It has long been debated whether GCs were made out of totally metal-free material or out of the same gas as the halo field stars  (progressively metal-enriched, during the halo chemical evolution). The former is  the ``classical" 
self-enrichment scenario (Cayrel 1986; Fall \& Rees 1985; Truran et al. 1991; 
Brown et al. 1991; Parmentier et al. 1999; Parmentier 2004), while the latter is a 
``pre-enrichment" scenario (e.g. Harris and Pudritz 1994). 

The classical self-enrichment scenario assumes that stars within the GCs made the totality of the metals (light and heavy) and it involves three stellar generations. A first generation of {\it exclusively } massive stars is clustered in the central regions of the progenitor cloud and their SNII explosions rapidly enrich the intra-cluster
gas in iron, $\alpha$- and heavy elements, but not in s-elements; the resulting [Ba/Eu] and [Sr/Ba] ratios should thus reflect pure r-process synthesis (Truran 1988). A second stellar generation is formed out of this gas, and a sub-class of rapidly evolving stars injects  into the ISM products of H-burning (and nothing else). The third generation formed from these ejecta has the same uniform heavy element content (in alpha- and heavier than Fe elements) as the second one, but a dispersion in its  light metal abundances. The long lived stars of the second and third generations are those observed today.

On the other hand the pre-enrichment scenario assumes that GCs were born with their current 
heavy metal  content and it invokes only two stellar generations: the first of them produces 
only the light metals which contaminate the gas of the second generation 
and builds up the observed anticorrelations 
(i.e.  they are 
equivalent to the second and third generations of the previous scenario).

The two scenarios differ in a key point: the slow neutron capture component of the heavy elements (s-component). In the first scenario, only the rapid neutron capture component (r-component), produced in SNII, may exist, whereas in the latter the s-component appears naturally (exactly as in the field stars of similar metallicity).
A recent high resolution and high S/N spectroscopic analysis of turnoff and early 
subgiant stars in three GCs spanning a large range in [Fe/H] values (James et al. 2004a, b) 
showed that the [Ba/Eu] and [Sr/Ba] ratios in GCs 
follow the trends found in field stars over the same metallicity range. Since Ba and Sr are s-elements, while Eu is an r-element, these findings
clearly indicate a similar and uniform enrichement 
by both r- and s-process nucleosynthesis in both populations (halo field stars and GCs).
Consequently, the ``classical" self-enrichment scenario is  disfavoured.

Another important result of the work of James et al. (2004a, b) is that no dispersion 
is found for s-element abundances within each individual GC (see also Armosky et al. 1994), 
nor any correlation with the O and Na patterns. 
This indicates that the stars responsible for the light element anticorrelations 
did not produce significant amounts of s-process elements. 

In view of the aforementioned observational constraints, one concludes that 
the GC heavy metals must come from pre-enrichment and that 
only light metals up to Al are produced in situ. 
{\it It is this scenario that will be called hereafter ``the self-enrichment scenario''}.

One key aspect of the problem is that the O-Na anticorrelation is found in all GCs 
where it has been looked for, independently of their characteristics, i.e., in disk 
and halo GCs spanning a large range in metallicity and in physical properties 
(such as horizontal branch morphology, total cluster mass, concentration and density; 
see Carretta 2006 and Carretta et al. 2006). 
It appears to be an intrinsic property of those stellar systems. 
Understanding the related self-enrichment process would obviously shed light 
on GC formation and on the physical processes which dominate the primordial phases 
of their evolution.

In particular one may try  to constrain the Initial Mass Function (IMF) of GCs 
in the framework of the self-enrichment scenario. 
Such an attempt was first made by Smith and Norris (1982) on the basis of the then available data
on C-N anticorrelation in NGC6752 and 47 Tuc. They found that the observed depletion of C in the cyanogen
enhanced stars of those clusters can be achieved only if the IMF of the first generation stars had
a slope $X<$-0.4 (where $X$=1.35 is the Salpeter value) and an upper mass limit $<$11 \ms.

In the present work we discuss the case of NGC~2808; it is the GC for which the largest sample of O and Na abundance determinations is available. On the basis of recent data, we derive the O/Na abundance distribution of the cluster (\S~2).
We explore two scenarios for the self-enrichment (\S~3), and in both cases we consider 
two possible classes of ``polluters" : (i) massive AGB stars (4 to 9~M$_{\odot}$) 
which undergo hot bottom burning and which are usually considered as the best culprits, 
and (ii) winds of rotating massive stars (WMS), which  offer 
an interesting alternative, as we argue in \S~5.
From the analysis of NGC~2808 data we constrain the 
IMF of the stars initially more massive than the ones surviving today,  
i.e., the GC IMF above 
0.8~M$_{\odot}$, and we discuss the implications 
for the amount of stellar residues (\S~4). In (\S~5) we discuss the various (often extreme) assumptions made in the course of our analysis and we argue that, in a realistic scenario, the IMF should be even flatter than the ones we derive in \S~4. We also discuss the advantages and shortcomings of the two classes of polluters, AGBs and WMS. In \S~6 we summarize our work.

\section{Abundance distribution of [O/Na] and self-enrichment in NGC~2808}

In the case of NGC~2808, the O-Na anticorrelation
was established for the first time by Carretta et al. (2004).
In an extended survey Carretta et al. (2006) detected both Na and O in 
91 RGB stars of the cluster and obtained upper limits 
for O abundances for another 36 giants with Na abundance determinations.
The spread in O and Na abundances is seen 
in stars of all luminosities along the RGB, 
pointing out  a primordial origin with no significant  
contribution from internal evolution of the stars (see also Carretta et al. 2003).

The data of Carretta et al. (2006) for the 91 stars where both O and Na 
are detected, are displayed in Fig.~\ref{ONaanticorrelation}.
It is clearly seen that the ratio [O/Fe]\footnote{In the usual notation of 
spectroscopists, [O/Fe]=log$_{10}$(O/Fe)$_*$-log$_{10}$(O/Fe)$_{\odot}$, 
where $*$ and $\odot$ denote stellar and solar values, respectively.} 
varies from $\sim$0.5 down to $\sim$--1, i.e. by a factor of $\sim$30,
while at the same time Na increases from [Na/Fe]$\sim$--0.2 to $\sim$0.8, 
i.e. by a factor of $\sim$10. 
Regarding the O-depletion levels, NGC~2808 is very similar to M13 
(Sneden et al. 2004); both clusters display the most severe variations
reported up to now.
Since the value of [O/Fe]=0.5 is typical of 
field halo stars (e.g. Spite et al. 2005), it is usually assumed that stars displaying 
that ratio (and correspondingly low [Na/Fe]) are those that maintain their 
original composition, the one 
of the proto-cluster gas. 
On the contrary, stars displaying the lowest [O/Fe] and highest [Na/Fe] 
have been severely contaminated by some external agent
(in the framework of the self-enrichment scenario). 
Stars with intermediate compositions  
are born from gas which was contaminated to a smaller extent.

Because of the anticorrelation, the ratio [O/Na] varies more than 
either [O/Fe] or [Na/Fe], from --1.55 to 0.55, and offers a larger basis 
for a statistical analysis of the abundance distribution within NGC~2808.
Such a distribution has already been presented in Carretta et al. (2006), 
but we adopt here a  different (and, in our opinion, more appropriate) 
one, obtained along the following  criteria:

\begin{figure}
\centering
\includegraphics[width=0.5\textwidth]{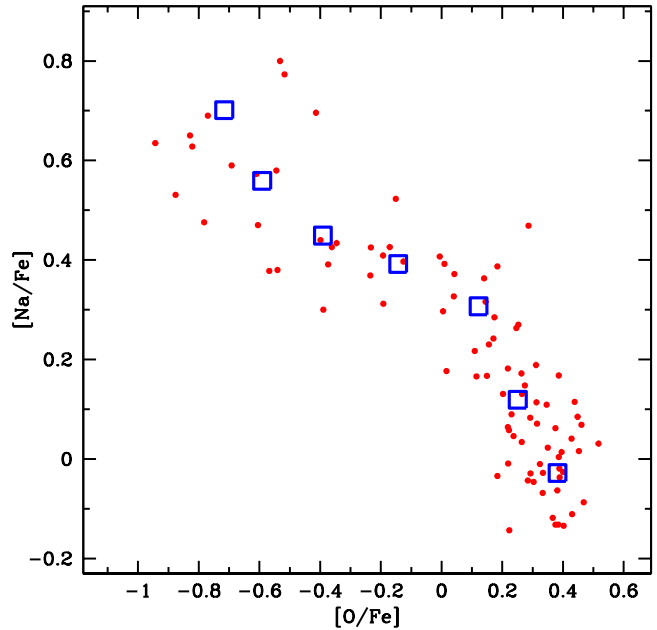}
\caption{Na vs O anticorrelation in NGC~2808. Data for a sample of 91 stars, 
with both Na and O detections, are from Carretta et al. (2006). 
{\it Open squares} indicate average abundances in the seven bins of 
the adopted O/Na distribution 
(see Fig.~\ref{histogramme} and \S~2).
\label{ONaanticorrelation}
} 
\end{figure}

The average error bars for [O/Fe] and [Na/Fe] in the 91 star sample of Carretta et al. 
(2006) are  $\bar{\sigma}_O$=0.077 dex 
and $\bar{\sigma}_{Na}$=0.098 dex, respectively. For each star, the  error
in the [O/Na] ratio $\sigma_{ONa}$=$\sqrt{\sigma^2_O + \sigma^2_{Na}}$ 
is calculated, and the
average is found to be  $\bar{\sigma}_{ONa}$=0.14. We consider then bins in [O/Na] of width
$\Delta$[O/Na]=0.30, i.e. slightly larger than twice the average error on that ratio. Provided the quoted error bars reflect reliably statistical errors, 
it can be reasonably argued that stars found in different bins have a 
trully different composition (in a statistically significant way) at least as far as O and Na are concerned.
The full range of the [O/Na] spectrum is covered by seven such bins, 
with values of [O/Na] ranging from --1.4 to 0.4. 
The average abundances of [O/Fe] and [Na/Fe] in the seven [O/Na] bins are 
shown in Fig.~\ref{ONaanticorrelation}.

\begin{figure}
\centering
\includegraphics[width=0.5\textwidth]{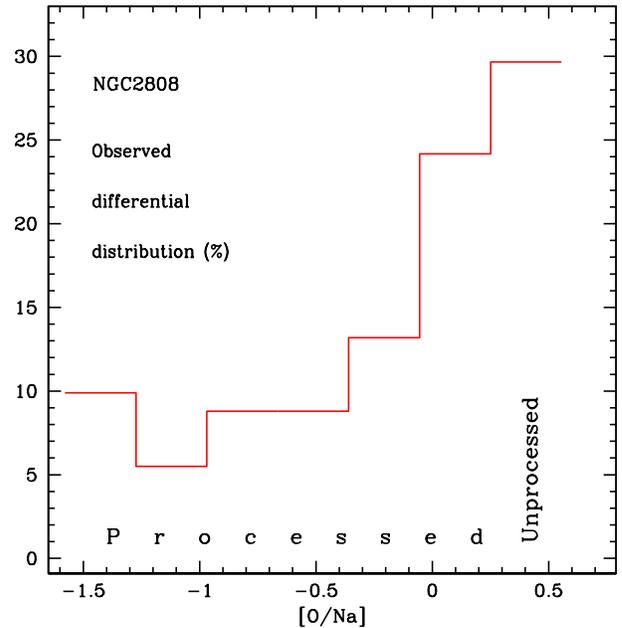}
\caption{Distribution histogram of stars of NGC~2808 (in percentages 
of total) as a function of [O/Na]; the distribution is 
calculated from data obtained by Caretta et al. (2006) (\S~2).
Stars in the rightmost bin (30\%) are presumably born with the original 
(unprocessed) composition of the cluster ([O/Na]=0.40) 
while the remaining 70\% have a composition processed to various degrees through H-burning. 
Scenario I (see text) assumes the processed composition to originate directly 
from the H-processed ejecta of AGB stars (4-9 \ms) or 
massive stars (10-100 \ms), born with the original composition. 
\label{histogramme}
}
\end{figure}

The resulting distribution histogram appears in 
Fig.~\ref{histogramme}. 
The number of stars in each [O/Na] bin has been multiplied by 100/91, i.e. 
an overall normalization to 100 stars has been applied, allowing one to read 
directly in the histogram percentages of the total.
Thus, 30\% of the stars in the rightmost bin have on average 
a composition of [O/Na]=0.40, i.e. the highest
O/Na ratio, which presumably reflects the original composition of the cluster. 
The remaining 70\% of the stars have a composition modified to various degrees by the
nucleosynthesis products of stars that evolved in relatively short timescales ($<$10$^8$ yr),
i.e. stars with initial mass M$>$4 \ms, according to the self-enrichment scenario
(e.g. Ventura et al. 2001).

\section{Polluting agents and pollution scenarios}

Usually, it is claimed that Hot Bottom Burning (HBB) in massive Asymptotic Giant Branch 
stars (AGBs) is responsible for the observed composition anomalies in GCs 
(Cottrell \& Da Costa 1981)\footnote{In their pioneer paper Cottrell \& Da Costa (1981) 
suggest that Na and Al enrichment in CN-strong stars of the GCs 47Tuc and NGC 6752 might be 
produced within intermediate mass stars ($\sim$ 5 - 10~\ms). However the invoked process
is neutron-captures on $^{22}$Ne and $^{25}$Mg within the thermal pulse (the neutrons 
being released by the $^{22}$Ne($\alpha$,n)$^{25}$Mg reaction; see Iben 1976).}.
Several custom-made detailed AGB models were recently computed in order to test 
the HBB hypothesis (Ventura et al. 2001, 2002; Denissenkov \& Herwig 2003; 
Karakas \& Lattanzio 2003; Herwig 2004a,b; Ventura \& D'Antona 2005a,b,c). 
Most of these studies concluded that the AGB pollution scenario suffers from 
severe drawbacks from the point of view of the nucleosynthesis 
(e.g., Fenner et al. 2004, Charbonnel 2005).
These difficulties stem from the subtle competition between HBB and 
third dredge-up (which contaminates the stellar envelope with the products of helium 
burning). The problems can be summarized as follows : 
(i) O is not depleted to the extent required by the observations, while Na
is over-produced or over-destroyed depending on the treatment of convection; 
(ii) Mg is produced while it should be destroyed, and
the Mg isotopic ratio is in conflict with the (rare) available data;
(iii) C+N+O does not remain constant in AGB processed material, in contrast to 
observations (Dickens et al. et al. 1991; Smith et al. 1996; Smith et al. 2005;
Ivans et al. 1999). 
Rotation-induced mixing makes the situation even worse (Decressin \& Charbonnel 2005;
Decressin et al. in preparation) by strongly enriching in O the envelope of intermediate-mass
stars, already during the second dredge-up event. 
Last but not least, it is sometimes  claimed that massive AGB stars are not expected to 
pollute the intracluster gas with s-process elements, in agreement with the observational 
requirements obtained by James et al. (2004)  and presented in \S~1; indeed, lower mass 
AGBs (1.5-3 \ms) are commonly considered to be the main source of s-elements in the Galaxy. 
However, s-process nucleosynthesis (from neutrons released by $^{22}$Ne($\alpha$,n)) may 
certainly occur if  temperatures in the thermal pulses  of massive AGB stars are sufficiently 
large (e.g. Goriely \& Siess 2005 and references therein). Very few quantitative studies 
have been devoted to that point, which clearly  deserves additional investigation. 
Until the question is settled, it should not be taken as granted that massive AGBs 
do not produce s-elements at all; in our opinion, this constitutes  another potential 
problem about their role as polluters of GCs. 

Although it is too early to discard the AGB hypothesis, we feel  that 
hydrogen burning in rotating massive stars 
may offer an interesting alternative (Decressin et al., in preparation). 
In fact, as we argue in \S~4, the latter alternative may be more 
attractive, as far as the constraints on the IMF are concerned; other advantages of that idea are qualitatively discussed in \S~5. 
In the following, we explore both possibilities 
for the mass range of the polluting stars, 
namely, we consider AGB stars of masses 4 - 9 \ms \ and massive stars of 
mass $>$10 \ms.

We  explore two scenarios for the self-enrichment, which differ by the degree of nuclear processing in the ejecta of the polluters  
and thus by the nature of the gas 
(i.e., pure stellar ejecta, or mixture of ejecta with gas of original composition) 
out of which the  low-mass stars still surviving today were formed.
In this section we present in detail the underlying assumptions
and we evaluate the amount of ejecta vs. the amount of long lived stars required
in each scenario. 
Then in \S~4 we derive the corresponding constraints on the IMF for each scenario, in the cases where either 
intermediate or massive stars are the polluters. 

\subsection{Scenario I}

The first scenario that we consider is the simpler one; it assumes that :

I-a) Stars with masses between 0.1 and 100 \ms \ are formed from gas of ``normal'' composition (i.e. the same composition as the one of field halo stars at that time, meaning for instance that [O/Na]=0.4) and with an IMF that has to be determined;

I-b) The ejecta of some (rapidly evolving) sub-class of those stars are 
polluted {\it to various degrees} by  internal nucleosynthesis (i.e. the [O/Na] ratio in the ejecta varies between -1.4 and 0.25), with an abundance distribution 
given by Fig.~\ref{histogramme}.
 
I-c) The ejecta are fully retained in the cluster and form 
directly (i.e., without being mixed with some original matter)
a second generation of stars with mass 
0.1-0.8 \ms, the same IMF as the one of the 1st generation low-mass stars, and with a composition 
obviously different from the original one (as far as the H-burning products 
are concerned);
 
I-d) Among the long-lived stars that we see today (with masses in the 
0.1-0.8 \ms \ range) in NGC~2808, 30\% are formed from the original gas and 70\% from the polluted ejecta.

We shall see that this ``Scenario I'' 
puts heavy constraints on the shape of the original IMF of the polluting stars, 
requiring it to be much flatter than the one observed today in Galactic and extragalactic 
stellar clusters. Indeed, it implies that the mass of the polluted ejecta should be about 
2.3 times larger (70/30=2.33) than the total mass of the first generation stars that survive 
today (of mass between 0.1 and 0.8 \ms). Moreover, if AGBs are assumed to be the polluting 
agents, we show that this scenario leads to the concomitant production of a large number 
of stellar remnants (mostly white dwarfs), the mass of which should be larger than the stellar 
mass of the cluster today. On the contrary,  if 
massive stars are assumed to be the polluting agents, the mass of stellar remnants is found to be considerably reduced. On the other hand, from the stellar nucleosynthesis point of view  
this scenario appears rather reasonable, since it invokes ejecta contaminated to various degrees (including very small ones) with the products of H-burning.

\subsection{Scenario II}

\begin{figure}
\centering
\includegraphics[width=0.5\textwidth]{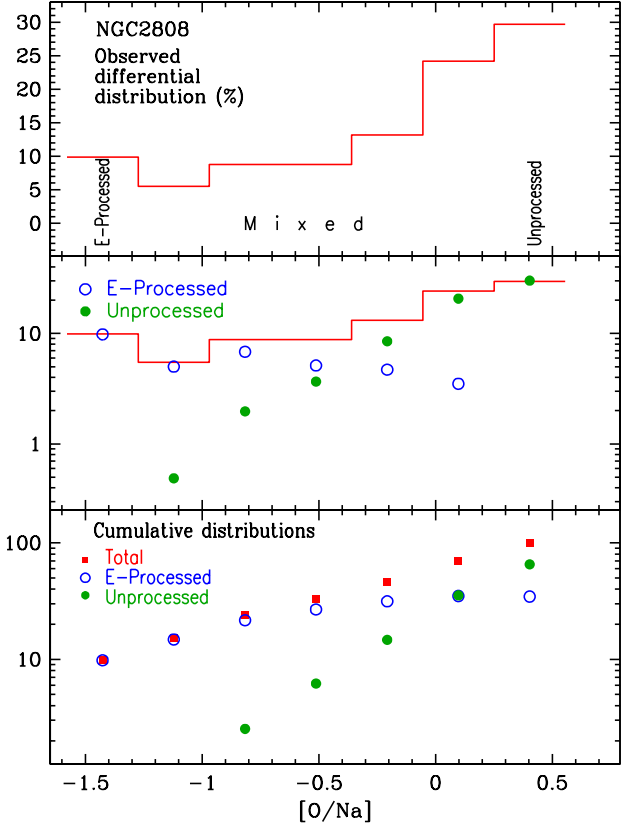}
\caption{{\it Top:} Distribution histogram of stars of NGC~2808 (in percentages 
of total) as a function of [O/Na] (same as in Fig.~\ref{histogramme}).
Stars in the rightmost bin (30\%) are presumably born with the original 
(unprocessed) composition of the cluster ([O/Na]=0.40) 
while the remaining 70\% have a 
contaminated composition. 
Scenario II (see text) assumes that the mixed composition is obtained by 
mixing extremely-processed material (E-processed, with [O/Na]=-1.40, leftmost bin) 
with unprocessed one, to various degrees, as explained in 
\S~3.2.
{\it Middle}: 
{\it open} and {\it filled} circles indicate the amount of extremely-processed 
($p_i$) and unprocessed  ($u_i$)
matter in each bin, respectively, while the histogram 
is the same as in the top panel (but in a logarithmic scale). 
{\it Bottom}: Cumulative distributions of total ({\it squares}), 
extremely-processed ({\it open circles}) and unprocessed matter 
({\it filled circles}), starting from the leftmost bin. Final values on the 
right are $P$=35\% for the extremely-processed and $U$=65\% for the 
unprocessed amounts of matter, respectively (see \S~3.2).
\label{scenario2}
} 
\end{figure}

Our second scenario assumes that:

II-a) Stars with masses between 0.1 and 100 \ms \ are formed from gas of 
``normal'' composition (i.e. the same composition as the one of field halo 
stars at that time, meaning for instance that [O/Na]=0.4) and with an IMF 
that has to be determined 
(this is identical to I-a);

II-b) The ejecta of some (rapidly evolving) sub-class of those stars are 
polluted {\it to an extreme degree} by  internal nucleosynthesis 
(i.e. the [O/Na] ratio in the ejecta is invariably -1.4);

II-c) The time-scale of the evolution of the polluters is shorter than the formation time-scale of the low-mass stars 
(0.1-0.8 \ms), i.e. shorter than about 100 Myr (Stahler \& Palla 2005). Then, the (extremely-processed) ejecta are mixed to various degrees with the gas of low-mass protostars of original composition, which are still forming.

II-d) Among the long-lived stars that we see today (with masses in the 
0.1-0.8 \ms \ range) in NGC~2808, 30\% are formed from the original gas 
(they have not been contaminated at all by the ejecta) and 70\% from 
various mixtures of the extremely-processed ejecta with the original gas.

Obviously, the starting and ending points of the two scenarios are identical (as they should, since they must both account for the observed abundance distribution in Fig.~\ref{histogramme}). 
They differ in the intermediate steps. Scenario I involves two clearly distinct stellar generations, the second of which is made exclusively by the nuclearly processed ejecta of the first one. 
Scenario II involves only one stellar generation, 
the low-mass stars of which 
(0.1-0.8 \ms) are contaminated {\it at formation } 
and to various degrees by the extremely-processed ejecta of their more massive and rapidly evolving sisters.

The assumption of extreme processing of the ejecta in Scenario II is neither mandatory (ejecta with various degrees of processing could also be assumed, as in Scenario I), nor easy to justify on physical grounds, as we shall discuss in \S~5. However, it is made here for two reasons: 

a) Combined with the abundance distribution diagram of NGC~2808 
(top panel of 
Fig. \ref{scenario2}) it allows one to evaluate in a well defined way the 
degree of mixing between processed and original material in the various [O/Na] 
bins (see middle and bottom panels of 
Fig.~\ref{scenario2} and next paragraphs).

b) Precisely because the composition of the ejecta is so extreme, 
less material of processed composition is required for the mixture. 
Consequently, the constraints on the IMF (from the required ratio between 
mass of ejecta and mass in low-mass stars of  0.1-0.8 \ms) are less severe than 
in Scenario I, as we show hereafter.

At this point, it should  be stressed that our assumption that only stars of 0.1 to 0.8 \ms \ 
are formed from the ejecta (points I-c and II-c) is also made only in order to minimize the constraint on the IMF of the polluters (since their ejecta are used in a most  efficient way, by  forming exclusively stars still visible today). If stars with M$>$0.8 \ms \ are assumed to be also formed from the ejecta, the required ejecta mass would be still larger and the corresponding IMF of the polluters even flatter than the ones we derive in the next section.

In order to evaluate the amount of ejecta with extremely-processed composition 
in the framework of Scenario II, we proceed as follows: assuming a mixture of 
unit mass of 
extremely-processed matter (mass fraction of a given element 
$X_{EProc}$) with $f$ parts of original matter (mass fraction $X_{Orig}$), the corresponding mass fraction $X_{Mixt}$ in the mixture will be:
\begin{equation}
X_{Mixt} \ = {{X_{EProc} + f X_{Orig}}\over{1+f}}
\end{equation}
For the ratio of two elements $a$ and $b$ one has then
\begin{equation}
\left({{X_a}\over{X_b}}\right)_{Mixt} \ = \ 
{{X_{a, EProc} + f X_{a,Orig}}\over{X_{b, EProc} + f X_{b,Orig}}}
\end{equation}
from which one may recover the 
dilution factor $f$ as
\begin{equation}
f \ = \ { {X_{a,EProc} \ -\ (X_a/X_b)_{Mixt} X_{b,EProc} }
\over{(X_a/X_b)_{Mixt}X_{b,Orig} \ - \ X_{a,Orig}}}
\end{equation}
as a function of the abundance ratio $(X_a/X_b)_{Mixt}$, provided the abundances of $a$ and $b$ in the original and the processed material are given.

In the framework of our Scenario II, we obtain the abundances of O and Na in the original and processed material by finding the corresponding  values in the rightmost and leftmost bins, respectively, of the adopted [O/Na] distribution. These values appear in 
Fig.~\ref{ONaanticorrelation} (where the averages [O/Fe] and [Na/Fe] in each of 
the adopted [O/Na] bins appear as open squares) and are:
$$[O/Fe]_{Orig} \ =0.38,$$
$$[Na/Fe]_{Orig}=-0.03,$$
$$[O/Fe]_{EProc}=-0.72,$$
$$[Na/Fe]_{EProc}=0.70. $$
The metallicity of NGC~2808 is [Fe/H]=--1.1 (Carretta et al. 2006), which allows one 
to find the absolute abundances of O and Na in the original and processed 
material. Turning all abundances into mass fractions, one may solve then 
Equ. 3 and find the 
dilution factor $f_i$ for each one of the [O/Na] bins 
($i$=1,...,7) in  Fig. \ref{scenario2}. 
Obviously, in the leftmost bin (material of extremely-processed composition 
exclusively) one has, by definition $f_1$=0, while in the rightmost bin 
(material of original composition only) very large values are found for 
$f_7$ ($f_7 \rightarrow \infty$). 

As mentioned in 
\S~2, the number of stars in each bin ($n_i$) represents also the percentage 
of material with (mixed) abundance ratio [O/Na] (since the total has been 
normalised to 100). The fraction
\begin{equation}
p_i \ = \ {{n_i}\over{1+f_i}}
\end{equation}
gives then the amount  of 
extremely-processed material (i.e. with [O/Na]=-1.4) in each bin, while 
the remaining
\begin{equation}
u_i \ = \ n_i \left(1-{{1}\over{1+f_i}}\right)
\end{equation}
represents the amount of unprocessed (original) material.
These amounts (also expressed as percentages of the total) appear in the middle panel of 
Fig.\ref{scenario2} (open circles for $p_i$ and filled ones for $u_i$).
Obviously, in the leftmost bin 
($f_1$=0) one has $p_1$=$n_1$, 
while in the rightmost bin 
$u_7$=$n_7$. By summing the amounts of the processed and unprocessed material 
over the seven bins, one obtains then the corresponding total amounts as
\begin{equation}
P \ = \ \Sigma \ p_i, \ \ \ U \ = \ \Sigma \ u_i
\end{equation}
Those quantities appear in the rightmost bin of the bottom panel of 
Fig. \ref{scenario2}. There is $P$=35\% of material with extremely-processed 
composition and $U$=65\% with original composition in the surviving stars of 
NGC~2808, in the framework of our Scenario II.

\begin{figure}
\centering
\includegraphics[angle=-90,width=0.5\textwidth]{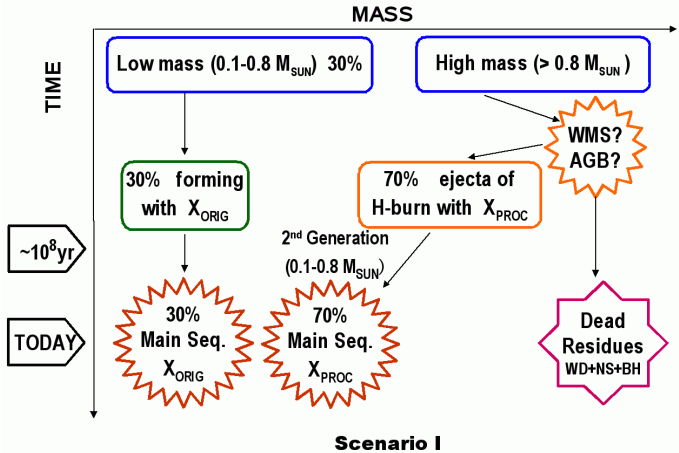}
\caption{Illustration of Scenario I 
with numbers corresponding to the analysis performed in \S~3.1. 
30\% of the currently surviving low-mass stars of NGC~2808 are born 
with the original composition. The heavier stars of that first generation 
(either in the AGB stage, or through the winds of massive stars - WMS)
eject 2.33 times more mass of varied composition, from which only low-mass 
stars (the remaining 70\%) is formed. This scenario puts relatively mild 
constraints on nucleosynthesis, but heavy ones on the 
initial mass function of the first stellar generation. {\it Percentages refer
to the total mass of low-mass stars of NGC~2808 (those still surviving today)}. 
\label{schemaS1}
} 
\end{figure}

\subsection{Summary}

\begin{figure}
\centering
\includegraphics[angle=-90,width=0.5\textwidth]{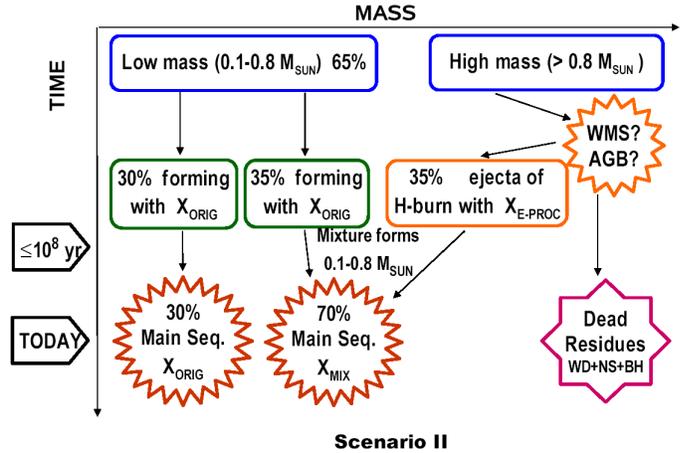}
\caption{Illustration of Scenario II,  
with numbers corresponding to the analysis performed in \S~3.2 
and Fig.~\ref{scenario2}. 65\% of the material of currently
 surviving low-mass stars of NGC~2808 is of  original composition ([O/Na]=0.40),
 and 35\% is of extremely-processed composition ([O/Na]=-1.40), ejected from
 either AGB stars or 
massive stars. The latter material is mixed with 35\% 
of the former to various degrees and forms low-mass stars (the currently surviving 70\% stars 
of mixed composition), while the remaining 30\% of the unprocessed material 
forms directly stars with the original composition.  This scenario puts relatively 
mild constraints on the initial mass function, but extreme ones on the 
nucleosynthesis, since it requires  all material ejected by the presumed 
sources (AGB or 
massive stars) to  have [O/Na]=--1.40. {\it Percentages refer to the total mass 
of low-mass stars of NGC~2808 (those still surviving today)}. 
\label{schemaS2}
} 
\end{figure}

The two scenarios considered here are 
sketched in Figures ~\ref{schemaS1} and \ref{schemaS2} respectively.
The numbers appearing on those figures are percentages of the final total mass in low-mass 
long-lived stars (0.1-0.8 \ms), derived from our  previous analysis (see also 
Fig. \ref{scenario2}). There is a big quantitative difference between the two scenarios, concerning the ratio of the processed ejecta mass to the mass of stars with original composition in the low-mass range. In Scenario I, this ratio is 2.33 
(=70/30), whereas in Scenario II it is much lower, only 
0.54 (=35/65). Of course, this difference stems from the 
different assumptions on the composition of the ejecta: processed to various degrees 
in Scenario I vs extremely-processed in Scenario II.

In the next section we explore the constraints imposed on the stellar IMF 
by the ratio mass(ejecta)/mass(low-mass stars) infered 
from our previous analysis.

\section{Constraining the IMF of the polluting stars in NGC~2808}

\subsection{The IMF in Globular Clusters}

Globular clusters provide, in principle, an ideal sample for an accurate evaluation of the 
stellar initial mass function (IMF) in the mass range $M \leq $0.8 \ms \ and down to the 
H-burning limit. Indeed, they are composed of stars with 
the same age, well determined distances and small variations in chemical composition 
(except for elements affected by H-burning) or in extinction within a cluster; 
moreover, the fraction of binary stars outside the cluster core  is less than 15\% 
(e.g. Rubenstein and Bailyn 1999) and affects little the determination of the luminosity 
function. Once the luminosity function is established, the existence of accurate stellar 
evolutionary models (reproducing well the observed color-magnitude diagrams of clusters with 
metalicities [Fe/H]$<$--1, e.g. Pulone et al. 1998) allows one to derive straightforwardly 
the IMF of low-mass stars.

The major potential problem in that enterprise is 
that formally, what we are measuring is the present-day mass function.
This could differ from the IMF because of 
the dynamical evolution of the cluster, which may be both of internal and external origin: internal, due to relaxation of the N-body system towards energy equipartition, by expelling less massive bodies towards the periphery of the cluster and accumulating massive ones in the center; and external, due to tidal interactions with the Galactic potential (or molecular clouds, 
other clusters etc.), which may result in the loss of the less massive bodies. Both effects lead to mass segregation of stars in space and time. Paresce \& de Marchi (2000, hereafter PdM00) found that the effects of mass segregation alter the stellar mass function both in the innermost regions (making it flatter than the IMF) and in those outside the half-light radius (making it steeper than the IMF); however, around the half-light radius, the effects of mass segregation are insignificant (see \S~4 and Fig. 4 in PdM00).

By using $HST$ measurements of the luminosity functions at (or just beyond) the half-light 
radius of a dozen globular clusters, PdM00 found that the corresponding IMFs have the shape of 
a lognormal distribution 
characterized by a peak mass $M_C$=0.33$\pm$0.03 \ms \ and a standard deviation 
$\sigma$=0.34$\pm$0.04\footnote{This agrees well with various studies 
(Zinnecker 1984, Adams \& Fatuzzo 1996, Elmegreen 1999) which show that one would expect 
a log-normal form for the IMF if the number of independent parameters governing the 
fragmentation of a proto-stellar cloud is large enough ($>$5).}.
Their measurements cover the mass range 0.09-0.7 \ms, but we shall 
assume here that the derived IMF shape is valid up to the  turn-off mass of $\sim$12 Gyr 
globular clusters, i.e. up to $\sim$0.8 \ms. We assume then that in the mass range of 
low-mass stars (0.1-0.8 \ms) the IMF is of the form

\begin{equation}
\Phi(M) \ = \ {{dN}\over{dM}} \ \propto \ {{1}\over{2 \ M}} \
\left({{log(M/M_C)}\over {\sigma}}\right)^2
\end{equation}
with values of $M_C$ and $\sigma$ given in the previous paragraph.
The IMF of more massive stars (living less than 
12 Gyr) is not observable in globular clusters, but it is measured
in young clusters, where it is found that it can well be
approximated by a power-law of the form

\begin{equation}
\Phi(M)  \ \propto \ M^{-(1+X)} 
\end{equation}
with $X$=1.35 being the ``classical" Salpeter (1955) value. As discussed in Kroupa (2002 
and references therein) the Salpeter value fits extremely well the available data over 
a wild range of cluster masses and in different galaxies, and with a dispersion 
of about 0.35, i.e. $X$=1.35$\pm$0.35.

In this work we assume that the IMF of the globular clusters was a composite one, given by 
Eqs. (7) and (8) for the mass ranges 
0.1-0.8 \ms \ and 0.8-100 \ms, respectively. We allow the value of $X$ to vary in the range -0.25 to 1.35 and we  constrain it using the results of the analysis in \S~3. All IMFs are normalised to
\begin{equation}
\int_{0.1}^{100} M \ \Phi(M) \ dM \ = \ 1
\end{equation}

\begin{figure}
\centering
\includegraphics[width=0.5\textwidth]{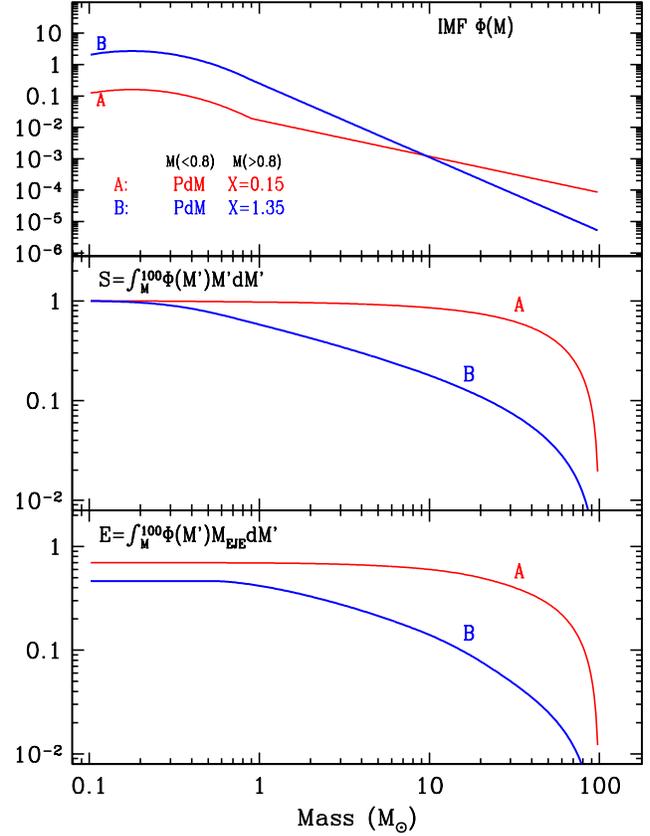}
\caption{{\it Top:} 
Examples of stellar IMFs studied in this work. They are composed of the Paresce \& de Marchi 
(2000, PdM) IMF of Eq. (7) in the range 0.1-0.8 \ms, and from the power-law IMF of Eq. (8) 
at $M>$0.8 \ms, with slope $X$=0.15 and 1.35 for IMF $A$ and $B$, respectively.  
All IMFs are normalised to $\int_{0.1}^{100} M \Phi(M) dM $=1. 
{\it Middle}: Corresponding total masses of the IMFs  
as a function of the lower integration limit $M$ in the integral  
$\int_M^{100} M' \Phi(M') dM' $. IMF $A$ has more mass in heavy stars 
(and, correspondingly, less in low-mass stars) than IMF $B$. {\it Bottom}:
 Corresponding Ejecta  mass fractions   as a function of the lower integration 
limit $M$ in the integral  $\int_M^{100} \varepsilon(M') \Phi(M') dM' $, 
where $\varepsilon(M)$ is the mass of the H-processed ejecta. Again, IMF $A$ has a much larger 
ejecta mass  fraction in massive stars than IMF $B$. 
\label{IMFexample}
} 
\end{figure}

Two of the adopted IMFs, with $X$=0.15 and 1.35, respectively,  are shown in 
Fig.\ref{IMFexample} (denoted with letters A, and B  respectively). 

\subsection{Mass of long lived stars, ejecta and residues}

A couple of other interesting quantities appear also in 
Fig.~\ref{IMFexample}. The mass fraction
 in the form of stars heavier than a mass $M$ is
\begin{equation}
S(M) \ = \ \int_M^{100}  M' \ \Phi(M') \  dM'
\end{equation} 
where, by definition, $S$(0.1)=1. Flat IMFs have a larger fraction of their mass in
 heavy stars, and a correspondingly smaller fraction in low-mass stars, than IMFs with 
a steep slope (see middle panel in 
Fig. 6). We also present (bottom panel in Fig. 6) 
the hereby defined {\it Ejecta mass fraction}
\begin{equation}
E(M) \ = \ \int_M^{100}  \varepsilon(M') \ \Phi(M') \  dM'
\end{equation} 
where $\varepsilon(M)$ is defined as {\it the part (in {\rm \ms}) of the star that is processed through H-burning only and is returned to the ISM}. Even with the help of detailed stellar models it is difficult to have precise values for 
$\varepsilon(M)$, since those values depend on several poorly known at present ingredients, like e.g. mixing criteria, mass loss rates, etc. In this work, we choose to deliberately maximise the amount of H-processed ejecta from each star, in order to obtain upper limits to the slope of the corresponding IMF. For that purpose, we adopt the following 
$\varepsilon(M)$ relations:
\begin{equation}
\varepsilon(M) \ = \ M \ - \ M_{Res}(M) \ \ \ \ (M<9 \ {\rm M}_{\odot})
\end{equation}
\begin{equation}
\varepsilon(M) \ = \ M \ - \ M_{He}(M) \ \ \ \  (M>9 \ {\rm M}_{\odot})
\end{equation} 
where $M_{Res}$ is the mass of the residue (white dwarf for M$<$9 \ms,  neutron star or black hole otherwise) and $M_{He}$ is the mass of the H-exhausted core of the star. This choice obviously maximizes the mass of the H-processed ejecta because it assumes that:

- In intermediate mass stars, all the mass outside the white dwarf remnant 
is processed exclusively through H-burning: in reality, part of that mass 
is  processed in the He-burning shell, but we assume here that this is 
negligible.

- In massive stars, all the mass outside the He-core is processed through 
H-burning; in reality, part of that mass is not processed at all (i.e. it is 
ejected with its original composition by the early stellar wind), while another 
part (close to the He-core) is contaminated by He-burning products; again, we 
assume those parts to be negligible, in order to maximize the mass of the 
H-processed ejecta.

\begin{figure}
\centering
\includegraphics[width=0.5\textwidth]{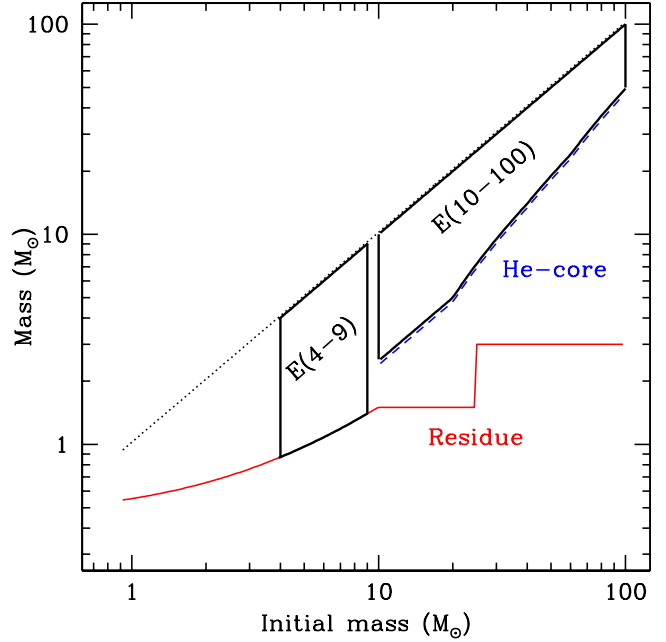}
\caption{Mass of He-core ({\it dashed curve}) for 
stars with M$>$9~M$_{\odot}$, and mass of stellar residues ({\it solid curve}) as a function of 
initial
stellar mass. The {\it dotted line} indicates the initial stellar mass, while the aereas within {\it thick curves} represent 
the H-processed ejecta that may contribute to the self-enrichment of globular clusters; the indicated amount is clearly an overestimate (see text). 
\label{residueejecta}
} 
\end{figure}

In Fig.~\ref{residueejecta} we display the adopted $\varepsilon(M)$ (from 
Eqs. 12 and 13), as well as $M_{Res}$, which is given from the relations:
$$
\hfill {\ \ \ \ \ 0.446 + 0.106 \ M \ \ \ (M<9 \ {\rm M}_{\odot}) }
$$
\begin{equation}
M_{Res}(M) \ = \ \ \   1.5 \  {\rm M}_{\odot} \  \ \ \ (10<M/{\rm M}_{\odot}<25)
\end{equation}
$$
\hfill {3 \ {\rm M}_{\odot} \ \ \ \ \ \ (M>25 \ {\rm M}_{\odot})}
$$

The white dwarf vs. initial mass relation for M$<$9 \ms \ is from Iben \& Tutukov  
(1984); it is assumed that stars in the 10-25 \ms \ range produce neutron 
stars of 1.5 \ms, while more massive ones produce black holes
of 3 \ms ( the mass of the neutron star or the black hole affects little the 
results discussed hereafter, as far as it remains below the mass of the 
He-core, which is a reasonable assumption).

The ejecta mass fraction $E(M)$ for the two IMFs $A$ and $B$ appear in the bottom panel of 
Fig.\ref{IMFexample}. Flat IMFs
have larger ejecta mass fractions than IMFs with steeper slope. 
Finally, in a similar way, the mass fraction locked in residues is calculated as
\begin{equation}
R(M) \ = \ \int_M^{100}  M_{Res}(M') \ \Phi(M') \  dM'
\end{equation}
From the defining equations (9) to (15) one may calculate any quantity of interest for a given IMF, 
e.g. the mass locked in low-mass stars (in the range 0.1-0.8 \ms) still surviving today is
\begin{equation}
S(0.1-0.8) \ = \ S(0.1) \ - \ S(0.8)
\end{equation}
while the ejecta mass fraction of stars in the 4 to 9 \ms \ range is
\begin{equation}
E(4-9) \ = \ E(4) \ - \ E(9)
\end{equation}
and similar relations hold for e.g. the mass of residues.

\begin{figure}
\centering
\includegraphics[width=0.5\textwidth]{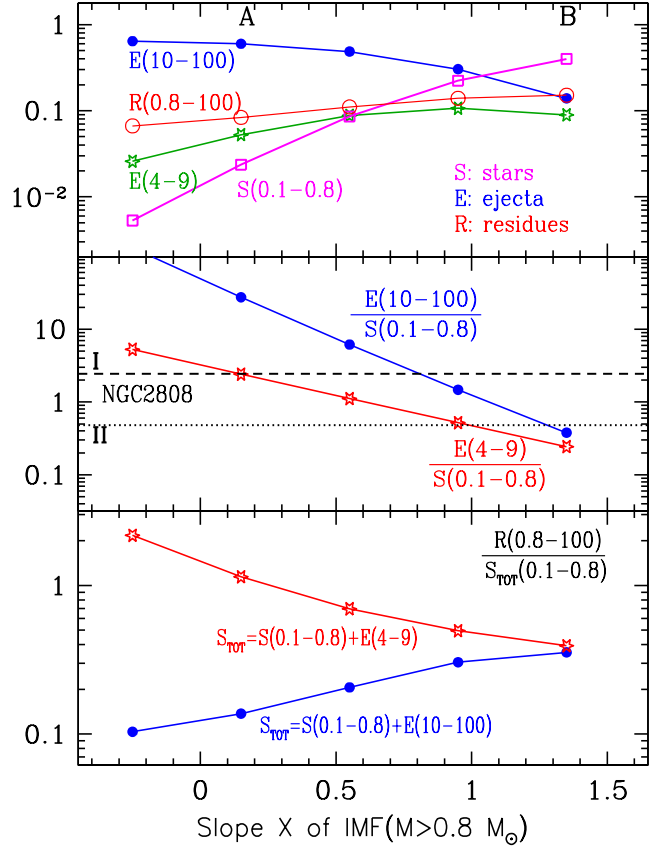}
\caption{{\it Top:} Mass fraction in low-mass stars surviviving today 
($S$(0.1-0.8 \ms), {\it open squares}), 
mass fraction $R$(0.8-100 \ms) 
in stellar residues (i.e. white dwarfs, neutron stars or black holes from stars
of initial mass M$>$0.8 \ms, {\it open circles}) and H-processed ejecta mass fractions $E$
from stars in the 4-9 \ms \ ({\it asterisks}) and 10-100 \ms \ ranges ({\it filled circles}), 
as a function of the slope $X$ of the high-mass part of the IMF; for the low-mass part, 
M$\leq$0.8 \ms, the PdM00 IMF
of Eq. (7) is always assumed.
Letters A and B  denote the results for the IMFs used in 
Fig. \ref{IMFexample}. 
{\it Middle}: Ratio of ejecta mass (of either 4-9 \ms \ or 10-100 \ms \ stars)
to the mass  of low-mass stars 
$S$(0.1-0.8) for each of the IMFs of our study. Note that 
$S$ represents stars (or material) of original (unprocessed) composition only. The two horizontal
lines correspond to the constraints imposed by the analysis of Sec. 2.2 and Figs. 2 and 3, according to
Scenarios I ({\it dashed}) and II ({\it dotted}), respectively. 
{\it Bottom}: Ratio of residue mass  to the total mass of low-mass stars surviving 
today, for each one of the studied IMFs. The total mass is the sum of the stars formed from 
original (unprocessed) material plus those formed from the ejecta of either AGBs (4-9 \ms) or
massive stars (10-100 \ms).
\label{IMFgeneral}
} 
\end{figure}

Four quantities of interest, calculated along the previous lines for 
all the IMFs studied here, are plotted in 
Fig.~\ref{IMFgeneral} (top panel) as a function of the slope $X$ of the $M>$0.8 \ms \ part 
of the IMF, which we wish to constrain. They are, respectively:

\begin{itemize}

\item{ The ejecta mass fraction $E$ from  massive stars (10 to 100 \ms); 
it is assumed that the envelopes of those stars are processed through 
H-burning at high enough temperatures, and are ejected by gently blowing 
stellar winds in the ISM, where they participate in the formation of new 
stars. }

\item { The ejecta mass fraction $E$ from Hot-Bottom Burning in massive AGB stars (of 4-9 \ms); these ejecta escape their stars at relatively low velocities, they are trapped in the cluster and form new stars. It is assumed here that the total envelope mass outside the 
Carbon-Oxygen white dwarf is processed that way.}

\item { The mass fraction $R$ in stellar residues (mostly white dwarfs), from all 
stars present in the original IMF that evolved until today (i.e. stars in 
the 0.8 to 100 \ms \ range).}

\item {The mass fraction $S$ in long-lived stars (0.1-0.8 \ms)}.

\end{itemize}

The first of these quantities increases when the slope of the high-mass part of the IMF decreases (when the IMF becomes flatter), while the opposite happens with the last quantity; the other two are relatively 
insensitive to the IMF slope.

\subsection{Constraints on the IMF of NGC~2808}

Armed with those numbers one may now explore the implications for NGC~2808 in the framework of the 
self-enrichment scenario. In the middle panel of 
Fig.~\ref{IMFgeneral} is plotted the ratio of the H-processed ejecta mass to 
the mass locked in low-mass stars of the original IMF (stars that still survive 
today) vs the slope $X$. 
Two cases are considered, namely that the ejecta originate either from 
massive stars 
(with mass 10-100 \ms) 
or from intermediate mass stars (AGBs of 4-9 \ms). Except for the steepest IMFs, the former ratio is always larger than the latter (since the ejecta mass of 10-100 \ms \ is larger than the ejecta mass of 4-9 \ms).
Of course, restraining  the mass range of 
massive stars that could contribute to the H-processed ejecta (assuming e.g. 
that only stars in the 30-100 \ms \ range develop sufficiently high 
temperatures for
the operation of Ne-Na or Mg-Al cycles) or assuming a smaller mass for the processed envelopes
of massive stars (instead of the Eq. 13 adopted here) would reduce accordingly the ejecta mass 
fraction of massive stars and would bring down the corresponding curves in 
Fig.~\ref{IMFgeneral} (top and middle panels). The same of course holds for the contribution 
of AGBs, which could well be smaller than assumed here
(for instance, if the mass of their H-processed ejecta is less than assumed in Eq. 12). 
Thus, the results presented in the middle panel of  Fig.~\ref{IMFgeneral} should clearly be 
considered only as {\it upper limits} to what AGBs or massive stars might do. 
In actual reality, all curves involving ejecta $E$ in the top and middle panels of 
Fig.~\ref{IMFgeneral} should be shifted downwards.

Those theoretical results are compared to the observationally inferred requirements 
for NGC~2808, for both Scenarios I and II in 
Fig.~\ref{IMFgeneral} (middle panel). As found in 
\S~3, Scenario I requires a ratio of H-processed ejecta to mass in 
low-mass stars with original composition of 70/30=2.33, while Scenario II requires 
a smaller ratio, of 35/65=0.54. As can be seen from 
Fig.~\ref{IMFgeneral} (middle panel), the requirement of Scenario I is 
satisfied only for slopes 
$X<$~0.8 in the case of massive stars and 
$X<$~0.15 for the massive AGBs. The requirement of Scenario II is satisfied for 
$X<$~1.25 in the former case and for $X<$~0.95 in the latter. 
IMFs with the standard slope $X$=1.35 fail to satisfy observational 
requirements by a factor of 2 in our Scenario II and by a factor of 10 in our Scenario I.

A diagram such as the one in the middle panel of Fig.~\ref{IMFgeneral} offers, in principle, 
a way to determine the slope of the high mass ($>$0.8 \ms) IMF of a globular cluster with 
a measured abundance distribution, as NGC~2808. However, this determination is made in the 
framework of models with extreme assumptions about both the polluter ejecta and their mixing 
with the pristine gas. In Sec. 5 we discuss these assumptions and we conclude that, in a 
realistic scenario, the IMF should be much flatter (i.e. $X$ should be much smaller) than 
determined in the middle panel of Fig.~\ref{IMFgeneral}. Still, it will be model dependent 
since the necessary ingredients (amount and composition of ejecta and degree of their mixing 
with the ISM) cannot be constrained independently at present.

At this point, it should be noted that
the IMF may affect critically the fate of a dense stellar cluster.  
Simulations of the dynamical evolution of GCs including the effects of tidal stripping 
and of mass loss from stellar evolution show indeed that a flat mass spectrum speeds 
up the cluster disruption (e.g., Joshi et al. 2001, Giersz 2001). 
This potential problem is alleviated in the present self-enrichment scenario, because the
mass lost by the polluters is assumed to be incorporated in low-mass stars of the cluster. 
Nevertheless the self-enrichment scenario needs to be tested quantitatively 
with realistic dynamical evolution models.

One way to constrain further the various scenarios is by searching for signatures left over 
in the form of stellar residues. This is the subject of the next section.

\subsection{Consequences for the amount of stellar residues}

In the bottom panel of Fig.~\ref{IMFgeneral} we explore the consequences of our different 
assumptions for the mass of ``dark objects'' in NGC~2808, namely residues of stars with 
initial mass 
M$>$0.8 \ms. Their mass (determined by the assumed IMF from Eq. (14) and given 
in Fig.~\ref{IMFgeneral}, top panel) is compared to the total mass of long-lived stars 
(M$\leq$0.8 \ms) of the cluster. The total mass is the sum of the original 
(unpolluted) low-mass stars $S(0.1-0.8)$ plus the polluted ones. If the 
polluted long-lived stars originate from the ejecta of 10-100 \ms \ stars 
(mass  $E(10-100)$), the sum increases considerably at low $X$ values and 
the ratio of residues to stars decreases in consequence. On the contrary, if 
the polluted long-lived stars originate from the ejecta of 4-9 \ms \ stars 
(mass $E(4-9)$), the sum decreases at low $X$ and the mass of the residues 
becomes comparable to the mass of active stars today (especially in case of 
Scenario I). Thus, {\it the mass ratio of residues to long lived stars} depends 
sensitively on the assumption made about the mass range of the polluters, 
especially for flat IMFs (such as those required to explain the abundance 
distribution of NGC~2808, and presumably of other clusters as well).

\begin{figure}
\centering
\includegraphics[width=0.5\textwidth]{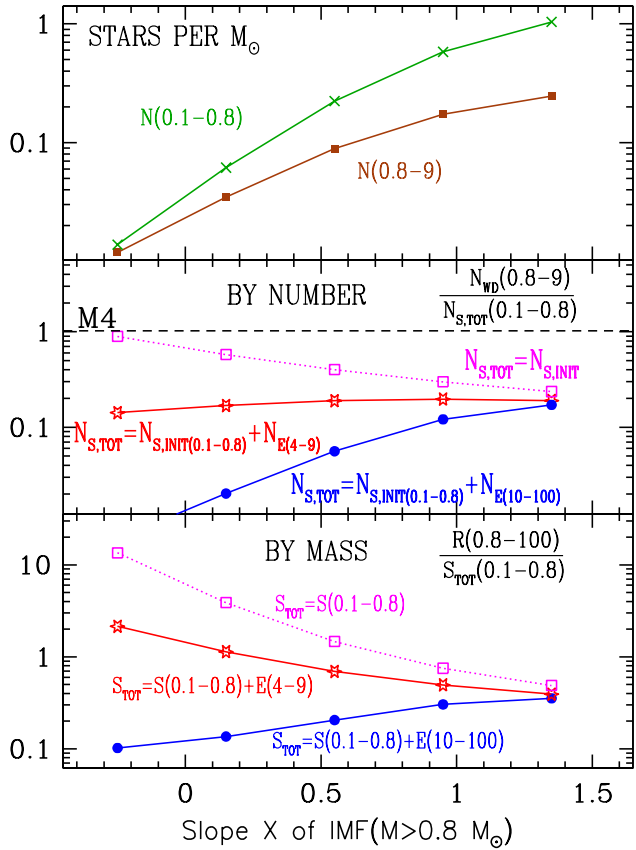}
\caption{{\it Top:} Number of stars per unit solar mass, as a function of the slope $X$ 
above the turn-off mass. The two curves corerspond to long-lived stars (0.1-0.8 \ms)
and to stars leaving behind white dwarfs (0.8-9 \ms \ initially).
{\it Middle}: Number ratio of white dwarfs to long-lived stars for the various IMFs studied in this work.
The two  {\it  solid curves} correspond to the cases where the polluters are AGBs (asterisks) and WMS (filled circles),
respectively (as in the middle and bottom panels of Fig. 8). 
The {\it dotted curve} is obtained by simply dividing the two curves of the top panel, i.e.
no creation of 0.1-0.8 \ms \ stars by the ejecta of polluters (self-enrichment) is considered. 
The horizontal {\it dashed line} represents the results of the study of Richer et al. (2002) for M4.
{\it Bottom}: Curves as in the middle panel, but for mass ratios instead of number ratios; also, 
the residues include not only white dwarfs (as in the middle panel) 
but also neutron stars and black holes. 
\label{Residues}
} 
\end{figure}

Unfortunately, such a ratio has not been determined in the case of NGC~2808. 
However, in the case of the cluster M4
(another mildly metal-poor GC with [Fe/H]$\sim$-1.08; Ivans et al. 2001), 
the {\it number ratio of white dwarfs to long lived 
stars} has been determined by Richer et al. (2002, hereafter RBF02). They used HST 
observations of that cluster to probe its luminosity function (LF) down to the H-burning 
limit ($\sim$0.09 \ms). 
Using stellar models of low-mass and low-metallicity stars (which still suffer from important 
uncertainties) they transformed the LF into a MF and concluded that the latter can be fitted 
by a power-law of slope $\alpha$=0.75 (or $X$=$\alpha$-1=--0.25 in the notation adopted here) 
in the mass range 0.09-0.65 \ms. In a subsequent work, however, de Marchi et al. (2004) argued 
that the use of more appropriate stellar models leads to a mass function similar to the one 
that PdM00 had determined for 12 GCs (Eq. 7). In the following we adopt this low-mass MF 
(the same used throughout our paper) also for M4.

By applying substantial completeness corrections to their sample, RBF02 derived the number 
of main sequence stars in the range 0.09 to 0.65 \ms \ and the number of white dwarfs from progenitor stars in the range 0.8 to 8 \ms:
they are, respectively, N$_{MS}$(0.09-0.65)=570 and N$_{WD}$=602. The number ratio of white dwarfs to long lived stars is thus $\sim$1. It should be stressed that these numbers are determined essentially by the applied completeness corrections, rather than by the observations (see Fig. 3, top panel, in RBF02).

RBF02 find that a single slope power-law IMF, with $\alpha$=1.05 (or $X$=0.05), i.e. the 
steepest slope compatible with their data, can accomodate this large ratio 
of N$_{WD}$/N$_{MS}$. They assume, of course, that both the observed MS stars and the WDs 
result from the original (first generation) IMF of the cluster, i.e they do not consider 
long-lived stars formed from the ejecta of the first generation. Since M4 is a cluster with 
observed O vs Na anticorrelation (Ivans et al. 1999, 2001), their IMF determination has a 
meaning only under the assumption that the observed anticorrelation is due to internal 
evolution and not to self-enrichment.
 
For a composite IMF, as the one adopted here (with PdM00 IMF below the turn-off mass of 
0.8 \ms  and a power law above it)  we find that, in the internal evolution scenario, 
a slope $X$=--0.3 above the turn-off mass is required to reproduce the 
derived N$_{WD}$/N$_{MS}\sim$1 (see dotted curve in the middle panel of Fig. 9). 

Somewhat surprisingly, in the self-enrichment scenario (solid curves in the middle and bottom 
panels of Fig. 9) such a high N$_{WD}$/N$_{MS}$ ratio is never obtained: values around 0.2 are 
obtained if the polluters are assumed to be AGBs and much lower ones if the polluters are 
massive stars. This may appear paradoxical, since at small IMF slopes and in the case of AGBs 
as polluters, residues dominate long-lived stars {\it by mass}, as clearly seen in Fig. 9 
(bottom panel). This apparent discrepancy is due to the fact that the ejecta mass of a e.g. 
4 \ms star can give rise to a dozen or so low mass stars ($\sim$0.3 \ms) of 2nd generation; 
thus, the resulting number ratio between WDs and long-lived stars is much lower than the 
corresponding mass ratio between WDs and ejecta.
Interestingly, in the case of internal evolution (implicitly adopted in RBF02), the residues 
should completely dominate by mass the long lived stars in M4, by a factor of 10, as can be 
seen in the bottom panel of Fig. 9 (dashed curve); this is due to the fact that each residue (white dwarf, neutron star or black hole) is several times more massive than a typical low-mass star.

The low number ratio of white dwarfs over low mass stars obtained in our self-enrichment 
scenarios (lower than 
derived in M4), does not necessarily point to a fatal flaw for those scenarios. It may well be that the ejecta mass (and the resulting number of second generation stars) is smaller than assumed here, in which case a ratio closer to the observed one will be obtained. Such a reduced ejecta mass is, in fact, quite plausible (see discussion in \S~5.1).

\subsection{Constraints from the Horizontal Branch in NGC2808}

A determination of the polluter IMF on the basis of star counts has been also attempted 
recently by D'Antona and Caloi (2004), who noted that AGB winds are helium-enriched 
(mainly as a consequence of the second dredge-up; Ventura et al. 2001). 
The second generation stars formed from AGB ejecta should then have a higher helium 
content than the first generation  stars, which were born with the primordial cluster
helium abundance. This difference in helium may alter substantially their advanced evolution.

D'Antona et al. (2002) evaluated the impact of a spread in helium within the low-mass 
star population on a GC color-magnitude diagram and discussed the evolutionary 
properties of GC stars in which variations of the initial helium are allowed. 
They showed that within reasonable limits (Y varying between 0.24 for the proto-cluster
abundance and 0.28 for the most polluted stars), the impact on the main sequence,
turnoff and RGB would be small and undetectable observationally. Also,
the horizontal branch luminosity appears to be only slightly affected.
As a consequence the age determination should be modified only in a limited way.
However the differences on the morphology of the horizontal branch would be
noticeable; in particular the He enrichment could play a role in the formation of blue tails. 

Following this idea, D'Antona \& Caloi (2004, hereafter DC04) studied in detail 
the morphology of the horizontal branch of NGC~2808 which is composed
of two well-separated regions : a red clump, and an extended blue tail starting 
from the blue side of the RR Lyrae gap and reaching below the cluster turn-off
(Bedin et al. 2000 and references therein). 
DC04 showed that the dichotomy of the horizontal branch of NGC 2808 could be 
reproduced by assuming different helium contents for the clump and blue HB stars, 
and argued that this unveils the signature of two separate events of star formation
in the early cluster evolution. Within their self-enrichment framework (which 
is similar, but not identical,  to our Scenario I), 
the red clump stars  belong to the first stellar generation formed with the 
original helium, 
while second-generation stars born directly from the ejecta of AGB stars of various 
masses (and thus with various He-enrichments)  populate the blue part of the HB. 
The lack of stars in the RR Lyrae region is attributed to the gap in helium 
between the two stellar generations \footnote{D'Antona et al. (2005) pushed further the analysis of the CMD of NGC~2808. 
They proposed that the main sequence width of this GC can be explained 
by the presence of three stellar generations born with different helium contents :
Y $\sim$ 0.24 (first generation born with proto-cluster abundance), Y $\sim$ 0.4 
(second generation, in chronological order, born from the winds of the most massive AGBs 
of the first generation), and Y $\sim$ 0.26 - 0.29 (third generation born 
from the ejecta from less massive AGBs of the first generation). In their scenario, 
late binary Type II supernova explosions are invoked to explain why 
there are no stars formed with Y between 0.29 and 0.4. We note that 
the third stellar generation should be contaminated by He-burning products (C in 
particular) and by s-elements ejected by low-mass AGBs which undergo third dredge-up 
but no HBB. Such abundance variations have not been discovered yet. See \S~1 
and our discussion in \S~5.2.1.}.

DC04 further explored the implications of their hypothesis by deriving the IMF 
of the first stellar generation based on star counts along the HB of NGC~2808. 
In order to explain the number of HB blue stars, they adopt a relation
between the progenitor AGB mass and the helium content of their ejecta (based on 
stellar models) and they derive a relatively flat IMF, with $X \sim 0$. This value 
is very close to the upper limit we derived in \S~4.3 for AGB stars being the polluters 
within Scenario I.

Helium enrichment is a natural outcome of AGB models. 
However the results of DC04 are extremely model-dependent; 
they rely in particular on a simple linear relation between the average helium 
content in the ejecta and the progenitor mass, require a fine-tuning of the
mass loss on the RGB, and depend on delicate transformations between the theoretical
and observational plane. Additionaly, other phenomena may intervene in the 
building of the color-magnitude diagram morphology of a GC (mass loss processes, 
stellar multiplicity, stellar rotation, dynamical effects) which should be ultimately 
taken into account in such an analysis (see e.g., Sosin et al. 1997, Castellani et al 2006 
and references therein).

Claims of supporting evidence for the presence of helium-rich populations in GCs were recently 
obtained through the detailed analysis of various color-magnitude diagram features 
(e.g., Lee et al. 2005 and references therein). 
This type of analysis brings new support to the self-enrichment hypothesis, but sheds no 
light to the nature of the polluters, since both AGBs and WMS are important He producers, 
as discussed in \S~5.
 
The results  presented and discussed in this section illustrate how deep counting observations of GC stars may be used to constrain the massive star 
(=polluter) IMF of the GCs, in the framework of the self-enrichment scenario. 
It is true that the  simplifying assumptions made in our work (upper mass limits for the IMFs
of the first - 100 \ms \ - and second - 0.8 \ms \ - stellar generations, same form for the low mass part of the IMFs in both stellar generations) are not unique, and a degenerate set of constraints would be obtained in a detailed analysis. However, with larger observational data sets available for a given cluster (including e.g. abundance distributions for several elements, number counts of white dwarfs and low mass stars - including those of the horizontal branch -, as well as  dynamical determination of the baryonic mass of the cluster etc.) and with improved theoretical understanding of stellar nucleosynthesis issues (e.g. amount and composition of ejecta) this degeneracy will ultimately be lifted.

\section{Discussion}

Today, there is not a unique, neither a well-defined, nor even a fully 
self-consistent scenario for the self-enrichment of globular clusters. 
The most commonly invoked one assumes that the H-processed ejecta of massive 
AGB stars of a first stellar generation (of unspecified, but presumably flatter 
than normal, IMF) form a second stellar generation (of equally unspecified IMF) 
and of appropriate chemical composition (i.e. in agreement with observed  O/Fe 
vs Na/Fe ratios in GC stars). Most of the theoretical effort up to now has been 
devoted in trying to reproduce the observed O vs Na or Mg vs Al relations
from the nucleosynthesis point of view (see references in Charbonnel 2005).

In this section, we discuss several aspects of the two self-enrichment 
Scenarios (I and II) which are sketched in 
\S~3 and which involve not only AGBs but also winds of
massive stars as alternative sources of the chemically polluted material.

\subsection{The mass and composition of the polluting ejecta}

We start by noting that any scenario of that kind necessarily involves assumptions about three quantities:

a) the IMF (mass limits and shape) of the first stellar generation and, when it exists,
of the second one also (there are two generations in our Scenario I but only one in our 
Scenario II);

b) the amount and composition of the stellar ejecta that contribute to the formation of the contaminated stars, as well as the mass range of the polluting stars;

c) the amount of unprocessed gas, which is mixed to various degrees with the processed ejecta (it is equal to zero in our Scenario I).

From those quantities, only the second can be evaluated, in principle, from stellar nucleosynthesis calculations; the other two can only be guessed at, or
constrained, by a combination of theory and observations. 

Since we lack, at present, the required input for item 
(b)\footnote{As mentioned in \S~3, the yields of massive AGB stars have been 
computed recently by several groups in the context of the self-enrichment hypothesis. 
Nevertheless, the predictions remain highly uncertain (mainly because they depend on the 
modelling of complex physical mechanisms such as HBB, third dredge-up, convection, mass loss, 
rotation, ...) and none of the present models does properly account for the O-Na 
and Mg-Al anticorrelations yet.
On the other hand, models of rotating massive stars with the appropriate initial 
composition will soon be available (Decressin et al. in preparation).}
we had to make here some assumptions. We assumed then that the 
{\it amount of H-processed ejecta} is the maximum possible from both potential sources, AGBs or massive stars (Eqs. 12, 13 and 
Fig.~\ref{residueejecta}). Obviously, any other assumption reducing the amount 
of the processed ejecta would result in IMFs even flatter than those found in 
\S~4 (all other things being kept the same). As discussed already in \S~4.2, a smaller amount of H-processed 
ejecta than assumed here is much more
plausible physically, for both AGB stars and  massive stars.

We also made two different 
assumptions about the {\it composition of the ejecta}, namely that they are processed to various degrees (Scenario I) or to an extreme degree
(Scenario II). Obviously, the latter assumption is utterly implausible, and it was used only to place absolute  lower limits
to the slope of the polluter IMF. Indeed, it is hard to conceive that the {\it whole envelope of all the polluters} (AGBs or massive stars)
 is processed to such an extreme degree ([O/Na]=-1.4); this may happen at some point of the evolution of some of the polluters, but it cannot
characterize to totality of the envelope evolution, nor the whole mass range of the polluters. 
Thus, the lower horizontal line of Scenario II (middle panel of Fig.~\ref{IMFgeneral}) 
is an extreme and unrealistic lower limit; the upper line
of Scenario I is more realistic in that respect. Still, observations suggest that it cannot be true, either.

Indeed, Pasquini et al. (2005) report the presence of Li  in 9 turn-off stars of the 
intermediate metallicity globular cluster NGC~6752 ([Fe/H]=--1.4), a cluster which displays 
the anticorrelation of O vs. Na. The Li abundance in NGC~6752 correlates with the one 
of O and anticorrelates with those of Na and N. The maximum Li abundance corresponds to stars 
with [O/Fe]$\sim$0.4, i.e. to stars formed from unprocessed material with composition similar 
to the one of field stars. It is no surprise then that the abundance of Li in those stars 
(log(Li/H)$\sim$--9.6) is at the level of the so-called ``Spite plateau"
observed in field halo stars (e.g., Charbonnel \& Primas 2005). 
On the other hand, the lowest detected Li values are about 3 times smaller and are found 
in stars with very low [O/Na]$\sim$--1.1 in NGC~6752.

The presence of so much Li (only within a factor of three below the 
plateau level)  in stars of modified composition in NGC~6752 clearly is at odds with our 
Scenario I: if such stars are made exclusively from the H-processed ejecta of some class 
of polluters (as assumed in our Scenario I), then no Li should be present there, in view 
of the nuclear fragility of that element. And even if one assumes that some fresh Li can be 
produced in H-burning (and ejected before subsequent destruction) through the 
``Cameron-Fowler (1971) mechanism", it is utterely implausible that the abundance of 
this new Li is almost as high as the ``Spite plateau". 
The most natural way to explain the presence of Li in the 
O-depleted stars of NGC~6752 is by assuming that those stars are formed from a mixture of 
nuclearly processed material  with material of normal  composition, as in our Scenario II
(Prantzos \& Charbonnel, in preparation).

Finally, we note that in both Scenarios I and II we assumed that, while the original IMF contained stars in the full mass range of 0.1 to 100 \ms, the ejecta of the polluters
formed only stars in the 0.1-0.8 \ms \ range, i.e. still alive  today. This assumption
has no physical justification and was only made in order to maximize the ratio of ejecta to live stars, in the same spirit with the other assumptions discussed in this section. Assuming that the ejecta formed stars in the full 0.1-100 \ms \ range would obviously lead to flatter IMFs in order to satisfy the observational requirements\footnote{Assuming that the second (``polluted") generation of stars contained its own massive and fast evolving ``polluters" (the ejecta of which would give rise to a third generation, and so on) would make the whole problem much more complex, and it would invalidate the analysis made here.}. 

We conclude then that both our Scenarios I and II are grossly underestimating ``reality", 
in the framework of the self-enrichment hypothesis for globular clusters. 
A realistic scenario should invoke both ejecta with varying composition (as in Scenario I, 
more realistic on nucleosynthesis grounds) {\it and} mixture of those ejecta with pristine gas 
(as in Scenario II, to account for Li observations); this implies that the horizontal 
``observational" 
lines in Fig.~\ref{IMFgeneral} (middle panel) should be replaced by a line lying above the 
one of Scenario I.  Also, a realistic scenario should 
consider an amount of ejecta considerably lower than  the 
(deliberate) overestimate we made in Eqs. (12) and (13); consequently,
the curves of Fig.~\ref{IMFgeneral}  (middle panel) should shift downwards, 
as discussed in \S~4.3. 
Overall, in a realistic scenario the slope $X$ of the polluter IMF should be much smaller 
than  derived in \S~4.3 and certainly much smaller than 0 in case the polluters are AGB stars. In other terms, an extremely 
flat IMF should be necessary to satisfy the constraints imposed by the abundance distribution 
of Fig.~\ref{histogramme}. 

In such a case, by extrapolating the trends obtained in the lower panel of 
Fig.~\ref{IMFgeneral} (towards the left) one sees  that:

- if the polluters are AGB stars, the mass of the cluster today should be completely dominated by residues (by a factor of several);

- on the contrary, if the polluters are massive stars, the total mass of the residues today would be much smaller than the mass of live stars in the cluster. 

With those implications in mind, we now take a closer look at the self-enrichment scenario, for each one of the two invoked classes of polluters.

\subsection{Polluters: AGBs vs. WMS}
 
\subsubsection{AGBs?}

According to a commonly held idea, the peculiar chemical composition of GC stars results 
from pollution by the ejecta of AGB stars, massive enough (4-9 \ms) as to develop HBB 
in their envelopes. Such stars evolve in time scales of $\sim$ 20 to 150 Myr and release 
their ejecta at velocities $v<30$ km/s (e.g., Loup et al. 1993), lower than the escape 
velocity from the cluster; the ejecta are trapped in the cluster and form new stars, 
either directly (our Scenario I) or indirectly, after mixing to various degrees with 
pristine material (our Scenario II, which accounts for the Li observations, 
as argued in Sec. 5.1).
 
In our opinion, this idea suffers from several drawbacks 
(in addition to the problematic nucleosynthesis predictions of the current models discussed 
in \S~3). In particular, it gives no satisfactory answer as to the roles of stars more 
massive (M$>$9 \ms) and less massive (M$<4$ \ms) than the presumed polluters.
 
1) Stars more massive than 9 \ms \ evolve on time scales lower than 20 Myr and are quite 
numerous (since a flat IMF is required in any case to account for the observed abundance 
distribution, as argued in this work). They release material processed through various 
nuclear burning stages, both through their winds (slow in the Main Sequence,
quite fast in the Wolf-Rayet stage) and through the final stellar explosion as 
supernovae. 
A commonly given answer as to the fate of their ejecta is that the fast moving supernova 
material ($v>$400 km/s even for the innermost SN regions) escapes the system and leaves 
no trace behind. On its way, it sweeps up the previously released wind ejecta, but also any 
ambient gas of pristine composition. Consequently, no original Li rich gas is left behind 
to get subsequently mixed  with the AGB ejecta. The second (polluted) stellar generation 
should then be formed exclusively with AGB ejecta, which implies that  AGBs should produce 
their own Li, at a level close to the Spite plateau. This is an extremely strong constraint 
on the nucleosynthesis  of AGB stars. 

Alternatively, one may argue that an insignificant amount of nuclearly processed ejecta 
escapes the supernovae, the vast majority being trapped within a black hole; this is an 
utterly implausible scenario, since neutron stars and pulsars are common in the Milky Way 
and there is no reason why massive stars in GCs would produce no such residues. 

Finally, one might also invoke a peculiar IMF of the first stellar generation in GCs, 
lacking massive stars altogether. Such an ad hoc assumption, however, is not supported by 
any theoretical arguments or numerical simulation results.

2) Stars in the mass range 2-4 \ms \ evolve on timescales of 150 to 1000 Myr 
 and release 
almost as much mass as those of 4-9 \ms (depending on the assumed IMF: for X=0.15 
we obtain $E$(2-4)=0.35 $E$(4-9)). Even if they do not develop HBB in their envelopes, 
they go through the third dredge-up phase; this brings He-burning products (in particular C) 
and s-elements to the surface, which are slowly released to the ISM. If new stars are formed 
from these ejecta, C and s-elements excesses should be observed in GC stars, which is not 
the case. One then has to assume that the ejecta of 2 to 4 \ms \ form no new stars, why those 
of their slightly more massive sisters manage to do it very efficiently. Perhaps, on time 
scales $>$100 Myr tidal interactions of the GCs with the Galactic potential are able to remove 
most of the cluster gas (as well as part of its stars); still some ejecta should be left 
behind to form stars with C and s-elements excess. Until a consistent dynamical scenario is 
developed to account  convincingly for that point, we consider it to be one 
of the drawbacks of the AGB scenario for the pollution of GCs.

\subsubsection{WMS?}

The idea that  WMS may be at the origin of some anomalies in the composition of GCs has been recently suggested by Norris (2004) and Maeder and Meynet (2006), in order to explain the blue main sequence of the cluster $\omega$ Centauri: the high helium content of the stars of that sequence could originate from WMS, producing a large helium/metal ratio.  Omega Cen is, however, a peculiar case,  showing a dispersion in its heavy metal content, unlike any other GC. If the polluters of GCs are indeed the winds of massive stars, the scenario could run along the following lines:

The winds of those stars, rich with H-burning products (especially if the stars are rotating),
are slowly released in the ambient ISM and they push it, opening circumstellar cavities;
they are finally mixed with pristine gas, still left over in the system after the formation of the first
stellar generation. By reaching the walls of those cavities, the shock waves of the subsequent SN explosions
induce the formation of new stars from this gas of mixed composition. The hot SN ejecta inside
the shock waves are expanding mainly along the cavities opened  up previously by the massive star winds, and they
finally find their way out of the cluster, escaping altogether the system. Thus, the stars
we now observe in the cluster are either of original composition, or of composition resulting from
the mixture of pristine gas with WMS. On longer timescales ($>$20 Myr) the wind ejecta
of lower mass stars (M$<$10 \ms) that do not explode, are slowly released in the cluster. However, they never
form new stars, since there is no external agent to trigger condensation of that gas and to 
induce star formation. As discussed before, tidal interactions ultimately strip that gas off the cluster.

The advantages of the WMS scenario are that: i) it may ``naturally'' account for the 
formation of the second stellar generation with mixed composition, which is triggered by 
the SN shock waves, and ii) it
may also explain why no trace of the He-burning products of the third dredge-up (which certainly occurs in 
low-mass AGBs) is seen today in GC stars: these products, as well as those of HBB in more massive AGBs, are
slowly released in a rarefied medium (since almost all of the gas of the system has been swept up or condensed
 previously, after the passage of the shock waves); as a result, they
never condense to form new stars (for lack of triggering agent) and they are ultimately removed from
the cluster through tidal interactions.

Of course, this novel scenario has also its potential drawbacks. It assumes that WMS mix thoroughly with
pristine gas and that the subsequent shock wave triggers an efficient star formation in the mixture, while
the SN ejecta escape the system. One may, in fact, imagine the opposite situation, where the shock 
waves disperse the gas, or the SN ejecta mix with that gas (at least partially). At the current stage of our knowledge, both alternatives are possible.

The WMS scenario sketched here is, admittedly, at a very qualitative level (but no more 
than the scenario involving AGBs). It should be substantiated at least by nucleosynthesis 
calculations demonstrating that WMS can indeed provide the chemical composition required
to explain the observations of abundances in GCs; such calculations are already performed 
and they will be presented elsewhere (Decressin et al. in preparation). 
We feel, however, that the WMS scenario opens very interesting perspectives for the 
understanding of the self-enrichment of GCs.

\section {Summary}

In this work we study the constraints on the IMF of stars responsible for the observed
abundance anomalies in GCs, in the framework of the so-called self-enrichment scenario.
We stress again that, at present, there is not a unique, nor a well-defined, nor even a 
self-consistent self-enrichment scenario for GCs. The parameter space is quite large and
 involves (at least) the IMF (mass limits and shape) of the polluter stars, the amount 
and composition of their  ejecta, as well as  the amount of unprocessed gas, which is mixed 
to various degrees with these ejecta in order to form the observed contaminated low-mass stars.

In view of the many uncertainties related to the self-enrichment scenario, we explore in 
some detail several possibilities, but our study is far from being exhaustive in that respect 
(\S~3). In particular, we explore two different types of self-enrichment scenarios, differing 
in the composition of the polluter ejecta, i.e. the degree to which the ejecta are 
contaminated by H-burning products:
in Scenario I, second generation stars are formed exclusively from ejecta processed 
at various degrees, while in Scenario II they are formed
from extremely processed ejecta mixed to various degrees with pristine material. 
Also, we explore two different possibilities for the polluters, namely AGB stars (4-9 \ms) 
and WMS ($>$10 \ms), since both of them can, in principle, release  in the ISM H-burning 
products at slow velocities. In each case we assume that the amount of ejecta is as large 
as possible, in order to constrain the polluter IMF from one side.

We adopt a composite IMF (\S~4.1) , with an observationally derived part in the mass range 
0.1-0.8 \ms (from PdM00) and a power-law for higher masses (i.e. for stars that are 
extinguished today); our aim is to constrain the slope $X$ of the power-law part of the IMF.
For a concrete illustration of our method we chose NGC~2808, the GC for which the largest 
sample of O and Na abundances is presently available. We derive the abundance distribution 
of O/Na in that cluster (\S~2) and find that $\sim$30\% of its stars have a pristine 
composition, while the remaining 70\% is contaminated to various degrees by H-burning products.

We then show how this contamination can be obtained quantitatively in the framework of our 
two self-enrichment scenarios and under the different assumptions made about the mass range 
of the polluters (AGBs vs WMS). We find that, in any case, the IMF of the polluters has 
to be flatter than the standard present-day IMF (i.e. with the Salpeter slope $X$=1.35); 
the trend is more pronounced in Scenario I than in Scenario II, and also more pronounced in
the case of AGBs than in the case of WMS (\S~4.3). However, due the many uncertainties 
involved, it is impossible to make a realistic quantitative prediction about the upper limit 
of the slope $X$; in fact, we argue in \S~5.1 that a realistic scenario would lead to an IMF 
much flatter even than those obtained in this work.

We note that an IMF with $X<$-0.4 (i.e. much flatter than obtained here) was found in the
work of Smith and Norris (1982) for GC 6752 and 47 TUC,  
in order to explain the observed C depletion in the framework of the self-enrichment scenario.
By using absolute abundances (and not only abundance ratios, as in our work)
Smith \& Norris (1982) were also able to place an upper limit of 11 \ms \ to the
IMF of the polluters. Clearly, such a work should be repeated with updated stellar yields
and a large base of spectroscopic data on as many elements as possible, in order to constrain
the polluter IMF and even the validity of the various self-enrichment scenarios.

We find that the amount of stellar residues (both  by mass and by number) present in the GC,
depends sensitively on the slope of the adopted IMF and on the mass range of the polluters; 
it provides then another potential 
discriminating test between the various possibilities. 
We illustrate this (\S~4.4) by comparing our results to the number ratio ($\sim$1)
of white dwarfs to low-mass stars 
inferred in the cluster M4. 
We find that our results fall short of producing such a large ratio, the discrepancy being 
more important in the case of WMS than in the case of AGBs. In fact, taken at face value, 
our results might be interpreted as a complete failure of the self-enrichment scenario 
(and as an argument for the internal evolution scenario). 
We argue, however, that this ``failure'' may be attributed  to the extreme assumptions
adopted here, and that  more realistic assumptions would produce results closer to the 
observations.

Finally, in \S~5 we discuss the various assumptions made in our study.
In particular, we comment on some problems associated (in our opinion) with AGBs, 
the most frequently invoked polluters of GCs (\S~5.2.1). We also develop in some detail, 
and for the first time, a possible scenario for the pollution of GCs by WMS, 
and we also invoke its own potential problems (\S~5.2.2). We conclude that WMS  offer
an appealing alternative to AGBs as polluter candidates of GCs.

Overall, our study shows that the self-enrichment scenario of GCs can be quantitatively tested,
through progress in both, observations (abundance distributions for several elements, number
counts of white dwarfs and low mass stars, dynamical determination of the baryonic mass of
a given cluster), and theory (improved understanding of
the amount and composition of the ejecta of the various candidate polluters). 

\begin{acknowledgements}
We are indebted to E. Carretta for providing us NGC~2808 data prior to publication. 
We wish to warmly thank G. Meynet for enlightening comments and a careful
reading of the manuscript, as well as T. Decressin, G. de Marchi, G. Meylan, 
and P. Molaro for useful discussions.
\end{acknowledgements}

{}

\end{document}